# Radial abundance gradients of 18 elements in Galactic open clusters from infrared MWM spectra

## A detailed analysis of 655 giants in 133 clusters

S. Bijavara Seshashayana,[1,2] H. Jönsson,[1] V. D'Orazi,[3,4] K. Cunha,[5,6] P. Frinchaboy,[7] J. M. Otto[7]

[1] Materials Science and Applied Mathematics, Malmö University, SE-205 06 Malmö, Sweden
e-mail: shilpa.bijavara-seshashayana@mau.se
[2] Nordic Optical Telescope, Rambla José Ana Fernández Pérez 7, ES-38711 Breña Baja, Spain
[3] Department of Physics, University of Rome Tor Vergata, via della Ricerca Scientifica 1, 00133 Rome, Italy
[4] INAF - Osservatorio Astronomico di Roma, via Frascati 33, 00078, Monte Porzio Catone, Italy
[5] Steward Observatory, University of Arizona, Tucson, AZ 85721, USA
[6] Observatório Nacional, Rua General José Cristino, 77, 20921-400 São Cristóvão, Rio de Janeiro, RJ, Brazil
[7] Department of Physics & Astronomy, Texas Christian University, Fort Worth, TX 76129, USA

**ABSTRACT**

*Context.* Open clusters are powerful tools for studying the Milky Way. While large spectroscopic surveys now provide spectra for many cluster members, automated pipelines and heterogeneous membership selections can introduce systematics and inflate apparent cluster scatter. Therefore, a homogeneous re-analysis with careful membership control and an explicit treatment of departures from Local Thermodynamic Equilibrium is valuable for establishing robust abundance gradients.

*Aims.* The aim is to derive precise Galactic radial abundance gradients for multiple elements using open cluster giants, and to investigate how these gradients depend on cluster age.

*Methods.* We re-analysed high-resolution infrared APOGEE Milky Way Mapper spectra from DR19 of the Sloan Digital Sky Survey for 655 open cluster members selected from Gaia data that satisfied strict quality cuts on signal-to-noise ratio. Stellar parameters and 18 elemental abundances were obtained using spectrum fitting with the Python version of Spectroscopy Made Easy, applying Non-Local Thermodynamic Equilibrium corrections for several key atomic species. Further quality control of the results was made by visual inspection of all fitted synthetic spectra.

*Results.* The metallicity of the clusters decreases with Galactocentric radius, following a global slope close to -0.06 dex/kpc. Beyond 10-11 kpc, there is modest flattening. In addition to the elements analysed in the Otto et al. (2026) study, we derive open-cluster gradients for V, Cu, Zn and Yb using APOGEE spectra. All elements exhibit negative [X/H] radial gradients of comparable magnitude, while [X/Fe] gradients are close to zero for the $\alpha$ and iron-peak groups. Several odd-Z and neutron capture species exhibit mild outward increases in [X/Fe], consistent with nucleosynthetic yields dependent on metallicity and/or delayed production channels for these elements. Old clusters have shallower gradients than young ones. However, the observed age dependence is influenced by the link between cluster age and galactocentric radius in the sample.

*Conclusions.* Compared with Otto et al. (2026), the present reanalysis yields generally smaller reported per-cluster abundance uncertainties and suggests distinct, element-dependent revisions to the inferred Galactic trends. While the global gradients align closely with the optical Gaia-ESO results by Magrini et al. (2023), we do not observe a straightforward flattening of the gradients with age. Our revised abundance scale reveals a shallower, broken radial metallicity gradient, providing a more robust observational benchmark for Galactic chemical-evolution models.

## 1. Introduction

Open clusters (OCs) serve as ideal laboratories for Galactic archaeology, as they enable detailed studies of the chemical and dynamical evolution of the Milky Way (MW; Friel 1995; Lada & Lada 2003). Furthermore, OCs have been utilised as testbeds for the calibration of stellar isochrones, the refinement of age-metallicity relations, and the constraint of models of Galactic chemical evolution (GCE) through homogeneous and well-characterised samples of stars. Numerous investigations, ranging from small targeted studies to large-scale surveys, have explored their properties. Large spectroscopic surveys such as the Apache Point Observatory Galactic Evolution Experiment (APOGEE; Majewski et al. 2017), GALactic Archaeology with HERMES (GALAH; De Silva et al. 2015), Gaia-ESO (Gilmore et al. 2012), and the Large Sky Area Multi-Object Fiber Spectroscopic Telescope (LAMOST; Deng et al. 2012) have made it possible to build upon earlier high-resolution studies by providing homogeneous and extensive datasets of stellar chemical abundances across the MW. The wealth of data provided by large spectroscopic surveys has been utilised in several large-scale studies (e.g., Frinchaboy et al. 2013; Cunha et al. 2016; Magrini et al. 2017; Donor et al. 2020; Spina et al. 2021; Myers et al. 2022; Magrini et al. 2023; Carbajo-Hijarrubia et al. 2024; Yang et al. 2025) to conduct extensive analyses of OCs across a broad range of ages and $R_{gc}$. Complementary to these large projects, coordinated efforts have focused on detailed analyses of stellar populations in OCs. For instance, the Stellar Population Astrophysics - Open Clusters (SPA-OC) series of papers comprises several studies that have examined giant and main-sequence stars using highly resolved optical and near-infrared spectroscopy. These studies have exploited the HARPS-N ($R \sim 115\,000$ in the optical) and GIANO-B ($R \sim 50\,000$ in the NIR) instruments (Bijavara Seshashayana et al. 2024a,b; Jian et al. 2024; Dal Ponte et al. 2025; Jian et al. 2025; Bijavara Seshashayana et al. 2025, and references therein). These studies provide valuable, high-precision, and high-resolution constraints on the chemical properties of individual clusters, complementing the broader, but typically lower-resolution, view offered by large spectroscopic surveys.

Large surveys offer statistically significant samples and homogeneous data; however, they frequently compromise precision for individual stars/clusters due to the need for automated analysis pipelines, varying data quality, and limited spectral resolution. Conversely, small-scale, targeted studies often offer higher-precision abundance determinations and detailed cluster characterisations. However, they are typically restricted to a limited number of clusters or elements, leading to potential selection biases and reduced Galactic coverage. Despite the substantial amount of data obtained, both large-scale and small-scale studies are subject to their own particular encumbrances. Due to this, studies of OCs have reported a broad dispersion in the measured metallicity gradients, with slopes ranging from approximately $-0.05 \pm 0.01$ dex/kpc (Reddy et al. 2016; Casamiquela et al. 2019) to about $-0.10 \pm 0.02$ dex/kpc (Jacobson et al. 2016). A significant number of these studies have focused primarily on the $\alpha$ and iron peak elements. In contrast, other chemically informative species, such as odd-Z and neutron capture elements trends have received less uniform constraints in OCs (Magrini et al. 2017; Donor et al. 2020; Myers et al. 2022; Yang et al. 2025; Spina et al. 2021, 2022). The majority of previous works have also concentrated on the inner and solar neighbourhood regions of the disk, leaving the outer Galaxy relatively less constrained (Magrini et al. 2023).

Among recent studies involving large samples, the latest paper from the OCCAM-project, Otto et al. (2026), provides the most relevant comparison for the present work, as they derive Galactic abundance trends from OCs observed within the Milky Way Mapper (MWM) survey. A major strength of their study is its use of a large, homogeneous survey dataset, providing a valuable benchmark for investigating abundance trends across the Galactic disc. However, the scale of such surveys naturally necessitates automated procedures that are optimised for consistency and efficiency across many stars. The present study therefore offers a complementary perspective, focusing on a carefully vetted subset of spectra and applying stricter quality control and a more detailed abundance analysis. In this way, the present work provides an independent re-examination of the trends reported by Otto et al. (2026), with the aim of refining the cluster mean abundances and assessing the robustness of the inferred Galactic gradients.

Classical Cepheids offer an alternative perspective on Galactic abundance gradients as they trace the young thin disk, providing insight into the current chemical composition of the Milky Way. In a recent study, Nunnari et al. (2026) analysed 401 Galactic Cepheids using high-resolution optical spectra and found that many abundance profiles are better described by nonlinear forms than by a single radial slope. This makes Cepheids especially useful for use as a reference for the young disk against which the age-resolved OC gradients can be compared. While Cepheids probe very young populations, OCs span a much broader age range and preserve information about disk evolution over several Gyr.

The present study aims at retaining the precision typical of small-scale studies while encompassing a sufficiently broad and homogeneous sample from the MWM catalog to investigate chemical abundance trends across the Galactic disk.

## 2. Data and sample

The data used in this work are based on observations from the Milky Way Mapper survey (MWM; Kollmeier et al. 2026), a component of the fifth phase of the Sloan Digital Sky Survey (SDSS Collaboration et al. 2025). MWM uses the two APOGEE instruments in New Mexico and Chile to collect data, and the MWM DR19 spectra that are the basis of this study includes the data from the previous APOGEE (Apache Point Observatory Galactic Evolution Experiment) and APOGEE-2 surveys, whose targeting strategy and design are described in (Frinchaboy et al. 2010; Zasowski et al. 2013, 2017; Beaton et al. 2021; Santana et al. 2021). The MWM data reduction is however slightly updated compared to the reduction made during the APOGEE surveys (Nidever et al. 2015).

For this study, we have constructed a new OC catalog that closely follows, and in several aspects improves upon, the MWM DR19 Value-Added Catalog of Open Cluster Stars (OCCAM DR19, Otto et al. 2026). In the present study, a cross-matching procedure was employed between all MWM DR19 spectra and the OC membership catalog of Cantat-Gaudin et al. (2020), based on Gaia DR2, and Hunt & Reffert (2023), based on Gaia EDR3. The retention of a star as a candidate cluster member was contingent on its membership probability exceeding 0.7 in either catalog, in conjunction with a Gaia RUWE value below 1.2. This approach differs from that of Otto et al. (2026), who primarily used Cantat-Gaudin et al. (2020) as the astrometric starting catalogue and only considered Hunt & Reffert (2023) as an alternative. As an additional external check on the selection of cluster members, we compared the present sample with the APOGEE DR17 catalogue of OC members from Guerço et al. (2025). However, since Guerço et al. (2025) is based on APOGEE DR17 and the present work uses MWM DR19 spectra, this comparison is used as a consistency check rather than as the primary basis for defining membership. After this, the sample was then reduced by applying the following quality cuts, based on the MWM parameters, to retain probable high-quality spectra for reanalysis: S/N $>$ 100, 3300 $<$ $T_{eff}$ $<$ 6000 K, 0.0 < $\log g$ < 3.7 dex, and -2.5 < [Fe/H] < 0.6 dex. No additional automatic cuts based on catalogue quality indicators or formal catalogue parameter values or uncertainties were imposed. Following the application of the quality cuts described above, each individual spectrum was visually inspected to eliminate stars exhibiting evidently problematic or unreliable spectra, thereby ensuring the retention of spectra suitable for the abundance analysis. Consequently, the outcome is not a direct reproduction of the OCCAM DR19 membership selection. After applying the combined Gaia-based membership requirement and spectroscopic quality cuts, and subsequent visual inspection, the final sample contains 655 spectra of 655 stars in 133 OCs. The difference in sample definition is therefore expected to affect the resulting cluster means and radial gradients, as well as the different abundance-analysis strategy discussed below. Table A.1 provides a detailed summary of the key properties of the analyzed stars and clusters, where cluster ages, distances, visual extinctions ($A_V$), and Galactocentric radius ($R_{gc}$) were adopted from Cantat-Gaudin et al. 2020. We adopted the cluster properties from Cantat-Gaudin et al. (2020) to ensure a homogeneous set of the above values across the entire sample, thereby maintaining consistency with the primary reference scale used in OCCAM DR19. In contrast, Hunt & Reffert (2023) was used to supplement the membership selection rather than to redefine these adopted cluster properties.

Fig A.1 illustrates the spatial arrangement of the cluster sample within the MW disk. The clusters are concentrated around the solar circle and extend mainly towards the outer disk, reaching $R_{gc}$ of approximately 21 kpc. The coverage in ($X_{GC}$, $Y_{GC}$) also illustrates that the sample probes a broad range of azimuth at intermediate and outer radii. This supports the interpretation of abundance trends with $R_{gc}$, age and [Fe/H] presented in subsequent sections. Although the final sample is smaller than that of Otto et al. (2026), it is of higher quality, as the more stringent selection process and manual vetting produce a cleaner and more reliable set of stars for the abundance analysis. To illustrate the visual-inspection step used in the quality control of the spectral fits, Fig. A.2 shows representative examples of accepted observed and synthetic spectra. The examples are shown for four cluster stars in a spectral region containing Mg I features. The observed spectra are well reproduced by the best-fitting synthetic spectra in both the local continuum and the line profiles.

## 3. Analysis

The Python version of Spectroscopy Made Easy (PySME v0.6.22; Piskunov & Valenti 2017; Wehrhahn et al. 2023; Jian et al. 2026) was employed to fit synthetic spectra to the observed spectra through $\chi^2$ minimization, both when determining the stellar parameters and elemental abundances. 1D Model Atmospheres in a Radiative and Convective Scheme (MARCS; Gustafsson et al. 2008) were used, and a non-local thermodynamic equilibrium (NLTE) analysis was applied to atomic lines of C, Na, Mg, Al, Si, S, Ca, K, Ti, Mn, Fe, and Cu (Amarsi et al. 2016, Amarsi et al. 2020, Mallinson et al. 2024, Caliskan et al. 2025, Amarsi et al. 2025). The atomic data used in this work are identical to those described in Montelius et al. (2022) and Nandakumar et al. 2023a,b, 2024a, to which we refer for further details. Briefly, the adopted line list is based on data from the VALD database (Piskunov et al. 1995; Kupka et al. 2000; Ryabchikova et al. 2015), with astrophysical adjustments of the $\log(gf)$ values for several atomic lines. Table B.1 shows a list of the atomic data used in the present work.

### 3.1. Stellar parameters and abundances

The determination of stellar parameters follows the method described in Bijavara Seshashayana et al. (2025), and we refer to that paper for an in-depth description. In short, $T_{eff}$, [Fe/H], $v_{mic}$, $v_{mac}$, and the abundances of C, N, Mg, Si, and Ti were simultaneously fitted using PySME using carefully selected OH, CO, and CN molecular lines, together with atomic lines of C, Mg, Si, Ti, and Fe. $\log g$ was not a free parameter but a "derived parameter" in PySME: it is calculated photometrically from the stellar age, mass, distance, $A_V$, and G magnitude, as the other parameters mentioned above are being optimized. In this procedure, the Gaia G magnitude, cluster distance, extinction, and age are kept fixed. The iterative aspect refers to the internal PySME optimisation: as $T_{eff}$ and [Fe/H] are updated during the $\chi^2$-minimization, the corresponding point on the adopted MIST isochrone is re-evaluated to obtain the stellar mass, and the bolometric correction and luminosity are updated consistently. $\log g$ is then recalculated from the standard relation between mass, luminosity, and $T_{eff}$. Thus, $\log g$ is iteratively recomputed *during* the optimisation. Likewise, [O/Fe] is treated as a PySME "derived parameter", following [Mg/Fe]. This approach is motivated by the shared nucleosynthetic origin of the lightest $\alpha$ elements, which are almost entirely produced in core-collapse supernovae (CC-SNe). In stars that are too warm to exhibit any OH lines, it is not possible to make direct measurements to constrain O. Therefore, aligning [O/Fe] with [Mg/Fe] helps to maintain a physically motivated chemical pattern (see, for example, Woosley & Weaver 1995; Nomoto et al. 2013; Kobayashi et al. 2020). After the fitting, all observed and synthetic spectra were visually compared and inspected in order to fine-tune the analysis for particular stars or remove the spectra from the sample completely in cases where the quality of the derived parameters could not be guaranteed, for example, in the case of the observed spectra showing very wide lines, due to fast rotation and/or unresolved binarity. In our analysis, we used continuum-normalised MWM spectra as input, but in the PySME analysis, any residual local mismatches in the continuum were addressed by applying a linear correction to the continuum within short selected spectral segments. The continuum-scaling coefficients were optimised alongside the synthetic-spectrum fitting process rather than through the re-normalisation of the entire spectrum. This local approach is particularly important for cool and metal-rich giant stars, as molecular blends can make continuum placement ambiguous.

Fig. C.1 shows the derived atmospheric parameters in the $T_{eff}$ to $\log g$ plane. Table D.1 shows the derived stellar parameters for the sample stars, alongside their quoted uncertainties. The average uncertainties for $T_{eff}$, [Fe/H], $v_{mic}$, and $v_{mac}$ are 20 K, 0.02 dex, 0.09 km/s, and 0.4 km/s, respectively. The random variation associated with the stellar parameters is reflected in these uncertainties, which can arise from several factors, including the S/N, possible difficulties in minimising the $\chi^2$ statistic, and, to some extent, continuum normalisation. Conversely, larger

systematic uncertainties arising from factors such as model atmospheres and atomic data remain more difficult to quantify.

The determination of stellar abundances also largely follow that of Bijavara Seshashayana et al. (2025). The fitting was performed separately for each element, yielding abundances for up to 18 elements per star. For a subset of spectra, abundance measurements could not be obtained for certain elements due to the absence of suitable diagnostic lines, insufficient S/N, or strong telluric contamination in the relevant wavelength regions.

The elemental abundance ratios [X/Fe] for all analysed species in the individual stars are listed in Tables D.2 and D.3, together with the formal fitting uncertainties returned by PySME. As with the atmospheric parameters, these formal uncertainties primarily reflect random effects, particularly those associated with the S/N. In contrast, the total error budget in stellar abundances is commonly dominated by systematic contributions, largely originating from uncertainties in the adopted stellar parameters and atomic data. The cluster mean abundances in Tables D.4-D.6 are accompanied by propagated formal uncertainties. For a cluster with $N$ valid stellar measurements, the uncertainty on the unweighted mean abundance is calculated as

$$\sigma_{\langle X\rangle} = \frac{\sqrt{\sum_{i=1}^{N}\sigma_i^2}}{N},$$

where $\sigma_i$ is the formal PySME uncertainty of the individual stellar abundance measurement. These values quantify the propagation of the formal fitting uncertainties into the cluster means. They typically range from 0.01 to 0.10 dex, with only a few cases approaching 0.20 dex. All mean cluster abundance measurements are shown as a function of [Fe/H], age, and $R_{gc}$ in Figs. 1-3. For comparison, Sinha et al. (2024) employed APOGEE DR17 and SDSS-V/MWM abundances to examine 26 OCs, placing $3\sigma$ upper limits on the intrinsic abundance scatter of OCs of less than 0.02 dex for most elements. However, weaker-line and neutron-capture species demonstrated more relaxed limits, reaching up to 0.2 dex.

## 4. Results and discussion

Cluster mean abundances are reported in Tables D.4-D.6, while the corresponding element trends as a function of metallicity, age, and $R_{gc}$ are shown in Figs. 1-3. The quoted uncertainties on the cluster means are propagated as formal uncertainties as described above. All abundance ratios are expressed relative to the solar reference values of Asplund et al. (2021). Radial gradients, quantified through linear fits to [X/H] as a function of $R_{gc}$, are summarized for the full sample, together with the results of Otto et al. (2026) and optical abundance gradients from Magrini et al. (2023), in Table 1.

### *4.1. Comparison of parameters and abundances*

A star-by-star comparison was conducted for the 655 stars common to both the present study and the SDSS-V MWM DR19 catalog. As both sets of measurements are derived from the same SDSS-V DR19 spectra, this comparison serves as an internal consistency check between two distinct analysis strategies, rather than as an external validation against an independent dataset. Within MWM DR19, stellar parameters and abundances are determined using the Astra[1] analysis framework, where ASPCAP provides a global spectroscopic solution followed by element-by-element abundance determinations (Mészáros et al. 2025).

By contrast, the present analysis uses PySME to determine $T_{eff}$, [Fe/H], $v_{mic}$ and $v_{mac}$ spectroscopically, while iteratively updating the $\log g$ value using photometry. During parameter determination, abundances of C, N, Mg, Si, and Ti are fitted explicitly, whereas [O/Fe] is tied to [Mg/Fe]. This treatment of O differs little from the ASPCAP global parameter step, in which the $\alpha$-elements are varied together and [O/Fe] and [Mg/Fe] are not fully independent parameters either. The main methodological differences, therefore, lie in the use of selected spectral windows, the photometric $\log g$ constraint, the adopted line list and NLTE treatment, and the individual visual inspection of all fitted spectra.

A comparison with the MWM catalogue reveals a high level of concordance between the two analyses, with the residual discrepancies primarily reflecting the distinct parameter constraints employed in each method. The most noticeable systematic difference is found in $\log g$, where the photometrically anchored values in the present analysis differ from the ASPCAP spectroscopic values by a median of -0.13 dex, with an r.m.s. scatter of 0.28 dex. In consideration of the diverse constraints imposed on $\log g$, the median offset of -0.13 dex is deemed to be modest in relation to the star-to-star scatter of 0.28 dex. The median offsets in $T_{eff}$ and [Fe/H] are also modest, with values of -20 K and -0.04 dex, and r.m.s. scatter of 84 K and 0.06 dex, respectively. $v_{mic}$ shows a small systematic difference, with the present values being approximately 0.04 km/s higher on average. Meanwhile, $v_{mac}$ shows a scatter of about 0.25 km/s. These quantities are not equivalent, and automated spectroscopic fits can include degeneracies between broadening terms and other fitted parameters. Therefore, a larger scatter is expected and does not necessarily indicate an inconsistency between the two analyses. However, this validation demonstrates the value of the present reanalysis: the use of photometric constraints on $\log g$, NLTE spectral synthesis, manual inspection of all fitted spectra, and a homogeneous line-by-line abundance analysis leads to a revised and more controlled abundance scale. These methodological differences are important because even modest systematic offsets in the stellar parameters can propagate into the cluster mean abundances and therefore affect the inferred radial abundance gradients. Appendix C shows the Kiel diagram of the analysed cluster stars and the comparison of the stellar parameters derived in this work with the OCCAM DR19 catalogue values.

Additionally, we specifically examined whether the lowest-$\log g$ stars show systematically lower metallicities relative to their host clusters, as reported in some previous studies (Spina et al. 2022; Carrera et al. 2022; Magrini et al. 2023). For the 13 stars with $\log g < 1.0$ dex, belonging to 11 clusters, the median residual relative to the mean cluster metallicity is -0.02 dex, with a median absolute deviation of 0.02 dex. Excluding clusters represented by a single analysed star yields a similar median residual of -0.03 dex, with a median absolute deviation of 0.03 dex. Although a few individual evolved stars are more metal-poor than their respective cluster means, these cases do not significantly affect the adopted cluster averages. We therefore find no evidence for a significant systematic metallicity underestimate among the lowest-$\log g$ stars in our sample.

[1] https://github.com/andycasey/astra

The comparison of Galactic trends reveals differences that extend beyond minor zero-point shifts. The metallicity gradient in the current sample is shallower, with d[Fe/H]/d$R_{gc}$ = -0.057 ± 0.004 dex/kpc, compared to -0.079 ± 0.006 dex/kpc reported by Otto et al. (2026). The present analysis also supports a broken radial profile, with a knee at $R_{gc} \simeq 10.6 \pm 0.8$ kpc and inner and outer slopes of -0.078 ± 0.009 and -0.041 ± 0.009 dex/kpc, respectively. Several element-specific trends, particularly for Ti, Na, Ce, and Nd, are also flatter in the present analysis. These findings are best interpreted as selective, element-dependent revisions to the DR19 cluster abundance trends, derived from a smaller but more rigorously vetted sample of 655 stars in 133 clusters, compared to the 1083 stars in 158 clusters analyzed by Otto et al. (2026).

Otto et al. (2026) focuses on bulk cluster chemistry and Galactic abundance gradients derived from the MWM analysis. Consequently, the most meaningful assesment is at the cluster level, where mean [Fe/H] and [X/H] values and their large-scale radial behavior can be directly contrasted. Differences are primarily due to membership definition, sample selection, and abundance analysis strategy within the shared DR19 framework. Figure 1 compares the cluster abundance ratios [X/Fe] from the present analysis with those from Otto et al. (2026). While both studies identify similar overall abundance trends, the present analysis yields more compact cluster abundance sequences and generally smaller reported per-cluster uncertainties. The difference is particularly evident in the high-uncertainty tails for Na, S, Cr, Co, Ce, and Nd. Certain elements, notably Ce and Nd, present challenges in APOGEE spectra due to their weak and frequently blended features. This difficulty has prompted specialized APOGEE abundance reanalyses, such as the BAWLAS catalogue (Hayes et al. 2022). BAWLAS rederived abundances for weak and blended APOGEE DR17 features, including Na, P, S, V, Cu, Ce, and Nd, utilizing the BACCHUS code, dedicated line-quality flagging, and upper-limit treatment. The significance of such specialized abundances for studies of Galactic gradients is further highlighted by Sales-Silva et al. (2026), who employed BAWLAS Ce and Nd abundances in analyses of chemical radial gradients in the inner Galaxy. The differences observed here for Ce and Nd compared to Otto et al. (2026) should be considered in light of the established sensitivity of these neutron-capture abundances to line selection, blending, continuum placement, and abundance determination methodology. Disregarding the larger scatter in [Ce/Fe] in Otto et al. (2026) as compared to our study, they also have systematically higher values of about 0.1 dex. This result is likely attributable to the systematic difference in $\log g$ between the analyses. The 0.13 dex higher $\log g$ in the MWM ASPCAP analysis yields a correspondingly higher derived [Ce/Fe] value, consistent with the observed difference (see Table 6 of Cunha et al. 2017).

Figure 5 compares the distributions of the reported per-cluster abundance uncertainties in the present analysis and in Otto et al. (2026), element by element. For the present analysis, each value entering a violin plot is the propagated formal uncertainty on an unweighted mean cluster abundance, calculated as $\sqrt{\sum_{i=1}^{N} \sigma_i^2}/N$ from the formal uncertainties of the individual stellar measurements. For Otto et al. (2026), the plotted values are the quoted per-cluster uncertainties reported in that analysis. For most elements, the present sample exhibits more compact distributions and shorter high-uncertainty tails than the Otto et al. (2026) sample, shown in blue. The contrast is particularly pronounced for Na, S, Cr, Co, Ce, and Nd, for which the Otto et al. (2026) distributions are broader and extend to larger reported uncertainties. Elements such as Mg, Al, Si, K, Ca, Ti, Mn, Fe, and Ni have more similar central values in the two samples, although the Otto et al. (2026) sample generally retains broader distributions. The principal difference is therefore the smaller number of clusters with large reported abundance uncertainties in the present analysis, rather than a uniform decrease in the median uncertainty for every element.

### 4.2. Radial Trends

The radial abundance gradients derived in this study are primarily compared with those reported by Otto et al. (2026), as these provide the closest reference, since both studies are based on the very same MWM DR19 spectra. This comparison highlights how two analyses based on the same DR19 material can yield different inferences regarding the slope, shape and element dependence of the Galactic abundance profile. Therefore, the scientific interest lies not only in whether the two studies agree point by point, but also in how the adopted membership selection, sample definition, and abundance methodology influence the inferred large-scale chemical structure of the disk.

For the metallicity gradient, Otto et al. (2026) report that the DR19 OC sample is adequately described by a single linear relation: d[Fe/H]/d$R_{gc}$ = -0.079 ± 0.006 dex/kpc and d[Fe/H]/d$R_{Guide}$ = -0.071 ± 0.005 dex/kpc for the full sample. Otto et al. (2026) also explored a bilinear parameterisation and found a break at a slightly larger radius, $R_{gc} \simeq 11.9$ kpc. However, they concluded that a single linear relation was statistically preferable for their sample. In the present sample, the corresponding linear fit (shown in Fig. 4) yields d[Fe/H]/d$R_{gc}$ = -0.057 ± 0.004 dex/kpc, which is shallower than the Otto et al. (2026) baseline over the radial range probed here. When a broken slope model is permitted, the preferred fit exhibits a knee at $R_{gc} \simeq 10.6 \pm 0.8$ kpc, with inner and outer slopes of -0.078 ± 0.009 and -0.041 ± 0.009 dex/kpc, respectively. The break radius was treated as a free parameter and estimated by minimizing the residual sum of squares over a grid of trial break positions. This was done while ensuring continuity of the segmented relation and requiring at least 10 clusters on each side of the break. Therefore, the main difference relative to Otto et al. (2026) is not only a modest shift in the overall slope, but also that the present analysis favors a broken radial profile, with the knee occurring at a smaller radius. The broken slope favored in the present analysis is consistent with the results obtained at very young ages (≈ 50-200 Myr) of the Classical Cepheids by Nunnari et al. (2026). Furthermore, using a large compilation of 1,879 OCs, Joshi et al. (2024) found that the Galactic radial metallicity gradient is better described by a broken linear relation. This has a steep inner-disc gradient of -0.070 ± 0.002 dex/kpc inside $R_{gc} \simeq 12.8$ kpc, and an almost flat outer-disc gradient of -0.005 ± 0.018 dex/kpc beyond this radius.

Another comparison can be made with the optical Gaia-ESO analysis in Magrini et al. (2023), which provides an independent reference based on a different wavelength regime and abundance analysis framework. Their OC sample covers a similar radial distance, approximately 6 $< R_{gc} <$ 21 kpc, and they report a metallicity gradient of $d[Fe/H]/dRR_{gc}$ = -0.054 ± 0.004 dex/kpc. This value is in excellent agreement with our result of -0.057 ± 0.004 dex/kpc. This suggests that the global metallicity gradient is robust against differences in spectral range, line diagnostics, and analysis method. However, an element-by-element comparison reveals modest discrepancies for certain species, indicating that individual abundance gradients are sensitive to sample definition, radial coverage, stellar parameter scale and the adopted abundance methodology.

As can be seen from Figs. 1-3, the radial behavior of the abundance ratios is strongly element-dependent. However, it is important to distinguish between [X/H] and [X/Fe]. In the present sample, all measured [X/H] ratios decline with radius, although the steepness of the decrease varies from element to element. The [Mg/H], [Si/H] and [S/H] ratios are all close to -0.05 dex/kpc, the [Ca/H] ratio is also clearly negative, and the [Ti/H] ratio is somewhat shallower at about -0.03 dex/kpc. In contrast, the corresponding [X/Fe] gradients for Mg, Si, S, Ca, and Ti are weak and mildly positive, remaining at only a few hundredths of a dex/kpc. The same behavior is evident in the Fe-peak group: their [X/H] slopes are uniformly negative, but most of the [X/Fe] trends are weak. V, Cr, Mn, Co, and Ni all remain close to zero, while Cu is only slightly positive. Although Zn shows a positive differential slope, the scatter between and within clusters is large.

The odd-Z and neutron-capture elements display the strongest element-specific variations. In [X/H], Na, Al, and K all decline with radius, with [Al/H] somewhat shallower. In [X/Fe], however, [Na/Fe] is approximately flat, whereas [Al/Fe] and [K/Fe] show mild positive gradients. A similar pattern is seen among the neutron-capture species. Both [Ce/H] and [Nd/H] decline by a few hundredths of a dex/kpc, while [Ce/Fe] and [Nd/Fe] are mildly positive rather than showing strong variation. Similar behavior was reported by Magrini et al. (2023) for Gaia-ESO OCs, although the positive [Nd/Fe] gradient in their study is steeper than that observed in the present analysis. Yb varies more strongly, with a negative [Yb/H] slope but a positive [Yb/Fe] trend. Overall, the radial behaviour suggests that the dominant large-scale signal in [X/H] remains metallicity-driven, whereas the corresponding [X/Fe] patterns are generally weak, nearly flat or only mildly positive, depending on the nucleosynthetic channel involved. This, in turn, suggests that much of the radial variation in the absolute abundances reflects the underlying metallicity gradient, while the differential trends with respect to Fe are smaller and more chemically selective.

Compared to the analysis presented in Otto et al. (2026), this study agrees on the broad disk abundance sequence, although certain element-specific inferences exhibit significant variation. For Mg, Si, and Ca, the differential behavior remains largely consistent between the two studies. The most pronounced disparities occur for elements whose trends are more sensitive to abundance precision, sample selection, or line-specific systematics, including Na, Ti, Ce, and Nd. In this analysis, the [X/Fe] trends are more clearly defined and, for several species, less extreme than those reported by Otto et al. (2026). The [X/H] distributions are also more distinct and less dispersed following stricter sample vetting. This is particularly relevant for elements such as Na, Ti, Ce, and Nd, for which the inferred radial behavior may be affected by both abundance uncertainties and the inclusion of less reliable cluster measurements. However, the present results do not imply a universal downward revision of every radial slope relative to Otto et al. (2026); rather, they suggest a more selective modification of the abundance pattern, with the largest revisions concentrated in a limited set of chemically informative species. Of course, direct one-to-one comparison remains limited to the elements reported in the OCCAM DR19 catalogue, and no strict external validation is possible for additional species derived here, such as V, Cu, Zn, and Yb.

Nunnari et al. (2026) report similar qualitative behavior as ours using classical Cepheids: [X/Fe] ratios remain largely unchanged across the thin disk, suggesting that many elements exhibit a coherent relationship with Fe. This finding lends weight to the idea that the dominant radial signal in [X/H] is primarily metallicity-driven, while the residual [X/Fe] trends reveal more subtle nucleosynthetic differences.

In summary, the comparison with Otto et al. (2026) suggests that the present analysis favors a somewhat different description of the metallicity profile, with possible evidence for flatter behavior in the outer disc beyond 10 kpc. The differential abundance trends for several key species are generally cleaner and less extreme than those reported by Otto et al. (2026). A comparison with Magrini et al. (2023) provides an additional external reference based on optical Gaia-ESO abundances. Despite the differences in wavelength range, sample selection, stellar parameter scale, and abundance methodology between the two studies, the present IR analysis yields broadly consistent gradients for several shared elements, and, in some cases, slightly flatter ones. This supports the view that radial abundance variations in the selected OC sample are relatively mild.

### *4.3. Age Trends*

Figure 6 presents the radial abundance gradients for four distinct age groups. For most elements from Fe to Cu, the gradients remain negative across all age bins, indicating that the general decrease in abundance with increasing $R_{gc}$ is preserved even when the sample is subdivided by age. However, the gradient strengths vary between elements, and the age dependence is not strictly monotonic. The youngest clusters typically exhibit steeper or comparably negative gradients for Fe, Mg, Ca, Na, Al, Mn, Co, and Ni, whereas the oldest clusters often show flatter gradients. This pattern is particularly evident among several $\alpha$-, odd-Z, and Fe-peak elements, although the differences between age bins are not always statistically significant.

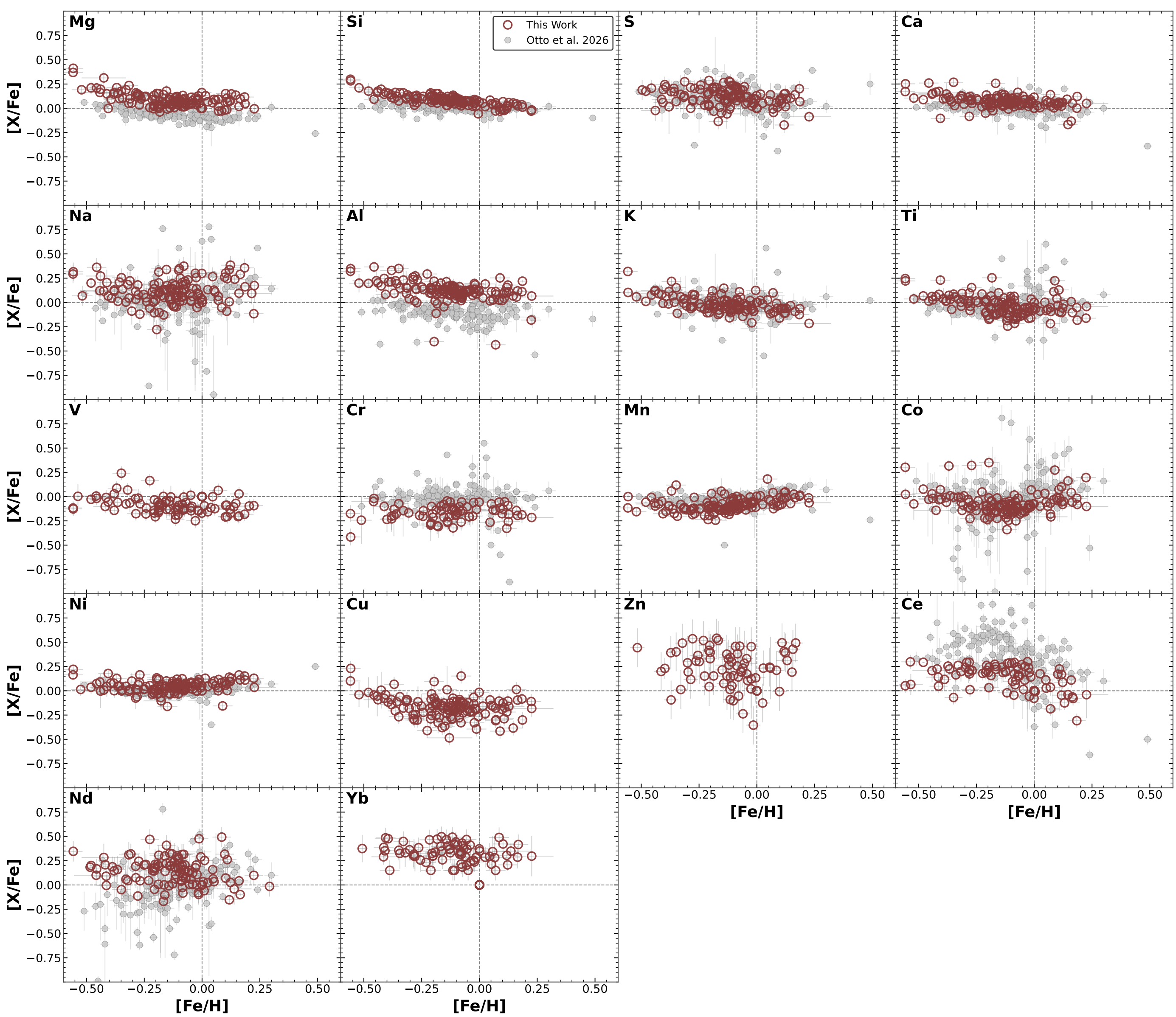

Fig. 1: Elemental abundance ratios [X/Fe] for our sample as a function of [Fe/H]. The red open circles are from the present work. The dark grey circles are from Otto et al. (2026).

In contrast, Ti, V, Zn, Ce, Nd, and Yb display greater variation and less consistent behaviour across age groups. Some age bins for these elements exhibit nearly flat gradients, while others show more negative slopes. This suggests that the inferred age dependence is influenced not only by chemical evolution, but also by the radial distribution of clusters, the sample size within each age bin, and the uncertainties in individual abundance measurements. Notably, neutron-capture elements do not demonstrate a simple progression from young to old populations, consistent with their more complex enrichment pathways and substantial inter-cluster abundance scatter.

Previous OC studies have arrived at differing conclusions regarding the age dependence of radial abundance gradients. Our result, that the youngest clusters frequently retain relatively steep gradients while older clusters tend to exhibit flatter gradients, contrasts with the Gaia-ESO OC results of Magrini et al. (2023), who reported flatter gradients for the youngest clusters. It also differs from Otto et al. (2026), who found no compelling evidence for significant age evolution in differential [X/Fe] gradients. Thus, a key finding of our analysis is that age-binned gradients are not simply a noisy manifestation of the global trend, but instead encode additional information about the chemical and dynamical evolution of the disc.

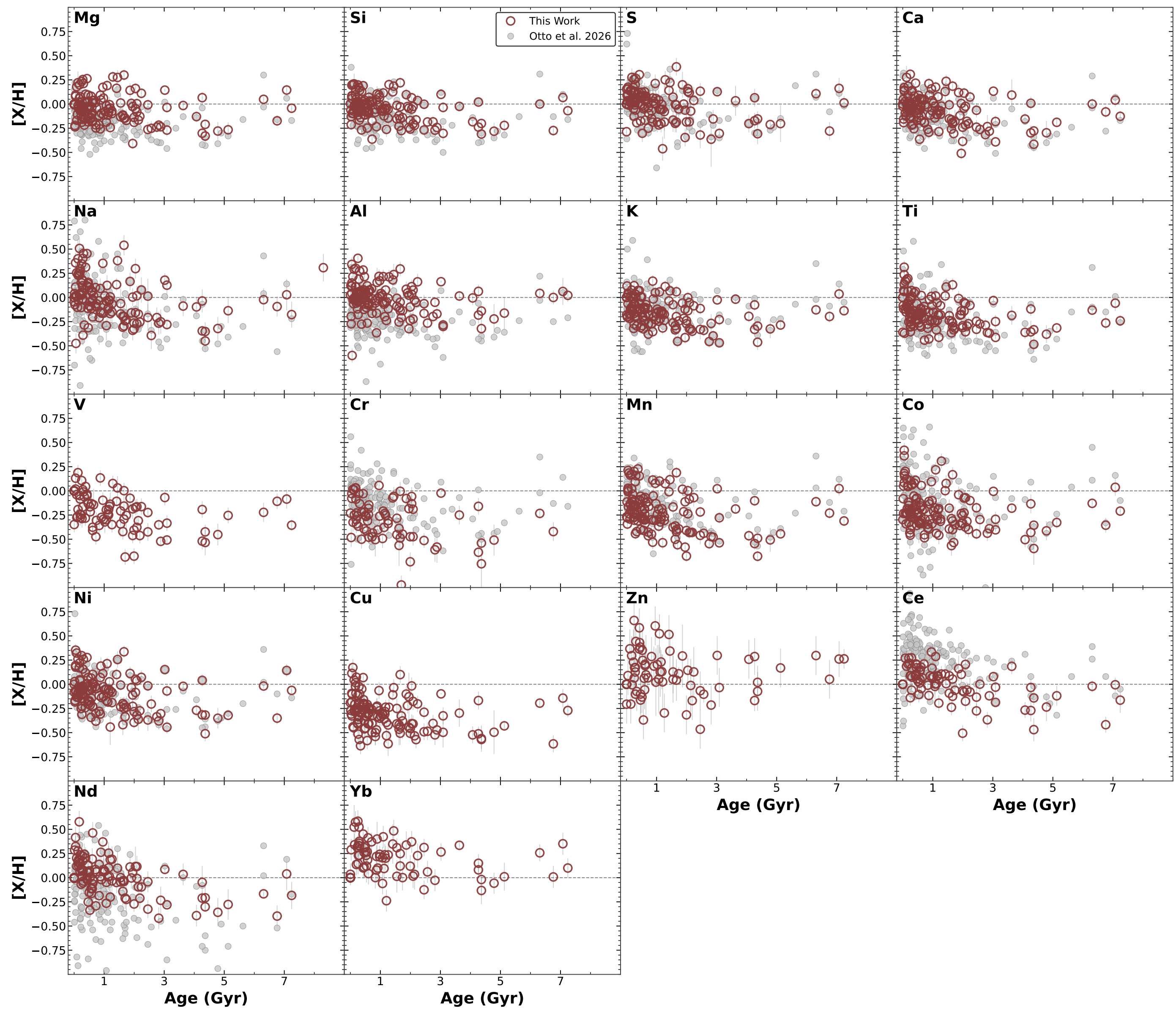

Fig. 2: Same as Figure 1 but here [X/H] values are plotted against age.

A useful benchmark for the neutron-capture elements is provided by Sales-Silva et al. (2022), who measured Ce abundances for 218 stars in 42 OCs using APOGEE DR16 spectra and the BACCHUS code. For clusters with $R_{gc} <$ 15 kpc, they found an overall negative [Ce/H] gradient of -0.070 ± 0.007 dex/kpc and a positive [Ce/Fe] gradient of 0.014 ± 0.007 dex/kpc. When subdivided by age, their [Ce/H] gradients were -0.033 ± 0.006, -0.033 ± 0.007, and -0.053 ± 0.018 dex/kpc for clusters younger than 1 Gyr, between 1 and 2 Gyr, and older than 2 Gyr, respectively. Corresponding [Ce/Fe] gradients were 0.018 ± 0.008, 0.027 ± 0.007, and 0.035 ± 0.007 dex/kpc, indicating only modest steepening of the positive [Ce/Fe] gradient with age. They further reported that younger clusters, particularly those younger than approximately 4 Gyr, generally have higher [Ce/Fe] and [Ce/$\alpha$] ratios than older clusters, which they interpreted as evidence for delayed and metallicity-dependent enrichment of the s-process by AGB stars. Similarly, Joshi et al. (2024) found that the age-metallicity relation cannot be described by a single trend, with a negative slope emerging for clusters older than approximately 240 Myr. This further highlights the sensitivity of inferred age trends to the adopted age range and sample selection.

The relatively steep gradients observed for the youngest clusters in this study can be understood if these clusters trace the current radial abundance structure of the disc, when, in contrast, older clusters are more likely to have experienced significant dynamical evolution, including radial migration and cluster survival effects. Such processes can redistribute clusters away from their birth radii and partially erase the original chemical gradient. Chemo-dynamical models, such as those of Minchev et al. (2014), demonstrate that radial migration and the time-dependent growth of the disc can substantially modify the observed relationship between age and radial abundance gradients, especially for older populations. Within this framework, the flatter gradients found for the oldest clusters are consistent with expectations that radial migration and dynamical heating weaken the connection between present-day location and birth abundance.

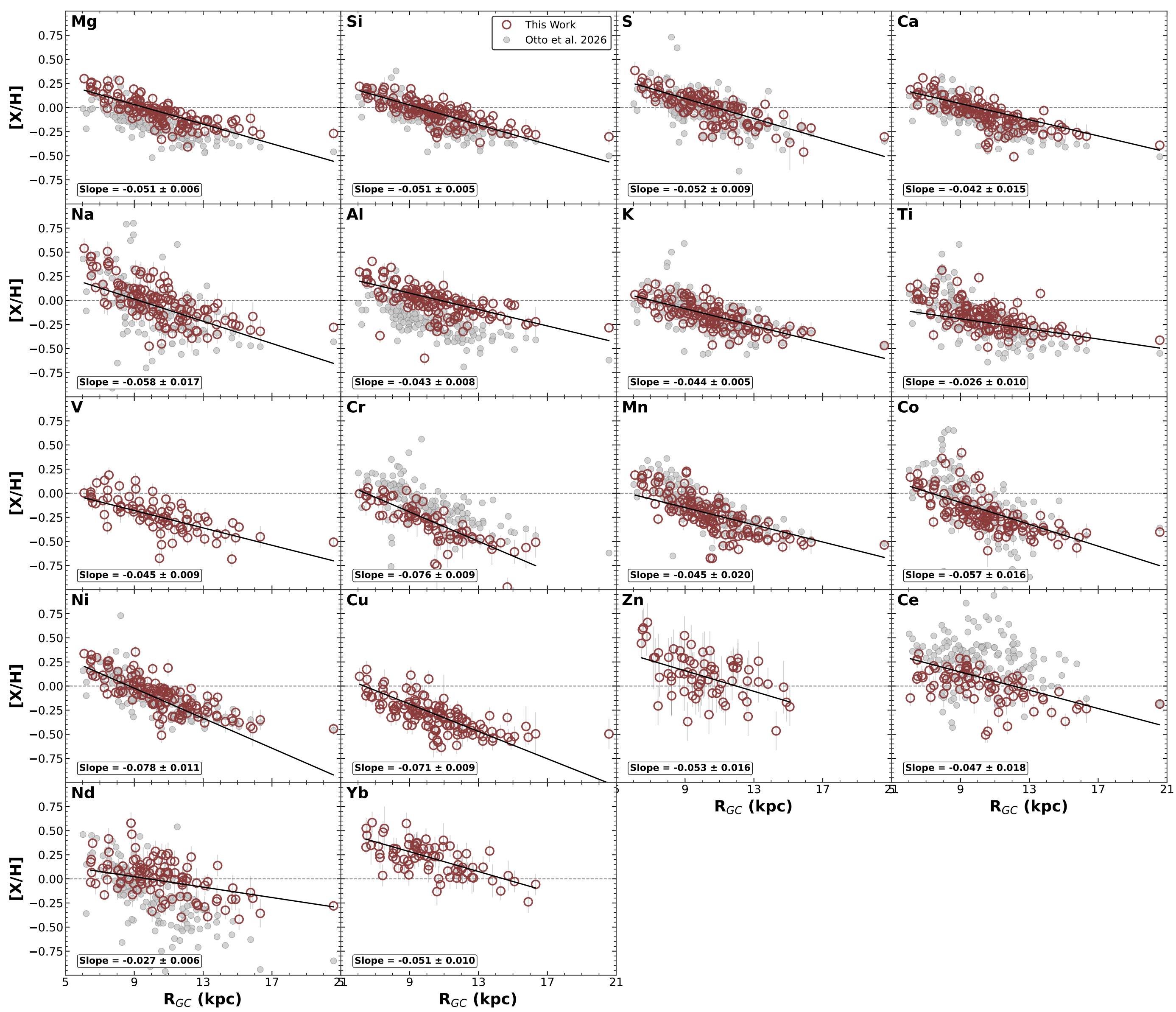

Fig. 3: Same as Figure 1 but here [X/H] values are plotted against $R_{gc}$.

### 4.4. Context within Galactic chemical-evolution models

The revised radial abundance gradients provide empirical constraints for GCE models. Where published predictions can be expressed in directly comparable observables, we make quantitative comparisons with the MWM OC results. In other cases, we restrict the discussion to a qualitative assessment because a rigorous model-data comparison would require the model tracks to be forward-modelled using the same radial coordinate, age selection, migration prescription, abundance uncertainties, and observational selection function as the present sample. Our global OC metallicity gradient, d[Fe/H]/d$R_{gc}$ = -0.057 ± 0.004 dex/kpc, is shallower than the OCCAM-DR19 value of -0.079 ± 0.006 dex/kpc reported by Otto et al. (2026), while agreeing closely with the Gaia-ESO OC gradient of -0.054 ± 0.004 dex/kpc measured by Magrini et al. (2023). Furthermore, the broken-line fit yields a steeper inner-disc gradient, $k_{in}$ = -0.078 ± 0.009 dex/kpc, and a flatter outer-disc gradient, $k_{out}$ = -0.041 ± 0.009 dex/kpc, with a transition radius at $R_{gc}$ = 10.59 ± 0.82 kpc. Within this parameterisation, the radial metallicity distribution is characterised by a comparatively steep decline in the inner disc followed by a more gradual decline in the outer disc. The age dependence of the gradients is also neither strictly monotonic nor identical for all elements. These results indicate that the present-day OC distribution reflects a combination of chemical enrichment, dynamical evolution,

| Element | This work | This work | OCCAM-DR19 Otto et al. (2026) | OCCAM-DR19 Otto et al. (2026) | Gaia-ESO Magrini et al. (2023) | Gaia-ESO Magrini et al. (2023) |
|---|---|---|---|---|---|---|
| | [X/H] | [X/Fe] | [X/H] | [X/Fe] | [X/H] | [X/Fe] |
| Fe | $-0.057 \pm 0.004$ | — | $-0.079 \pm 0.006$ | — | $-0.054 \pm 0.004$ | — |
| Mg | $-0.051 \pm 0.006$ | $0.009 \pm 0.002$ | $-0.048 \pm 0.040$ | $0.010 \pm 0.006$ | $-0.058 \pm 0.006$ | $0.009 \pm 0.003$ |
| Si | $-0.051 \pm 0.005$ | $0.015 \pm 0.002$ | $-0.057 \pm 0.040$ | $0.003 \pm 0.006$ | $-0.039 \pm 0.004$ | $0.002 \pm 0.002$ |
| S | $-0.052 \pm 0.009$ | $0.009 \pm 0.002$ | $-0.048 \pm 0.040$ | $0.025 \pm 0.010$ | — | — |
| Ca | $-0.042 \pm 0.015$ | $0.007 \pm 0.002$ | $-0.056 \pm 0.040$ | $0.003 \pm 0.004$ | $-0.034 \pm 0.003$ | $0.018 \pm 0.003$ |
| Na | $-0.058 \pm 0.017$ | $0.000 \pm 0.004$ | $-0.064 \pm 0.040$ | $-0.027 \pm 0.020$ | $-0.061 \pm 0.005$ | $0.003 \pm 0.002$ |
| Al | $-0.043 \pm 0.008$ | $0.016 \pm 0.004$ | $-0.052 \pm 0.040$ | $0.008 \pm 0.010$ | $-0.046 \pm 0.007$ | $0.012 \pm 0.004$ |
| K | $-0.044 \pm 0.005$ | $0.008 \pm 0.002$ | $-0.048 \pm 0.040$ | $0.015 \pm 0.010$ | — | — |
| Ti | $-0.026 \pm 0.010$ | $0.011 \pm 0.003$ | $-0.068 \pm 0.040$ | $-0.016 \pm 0.010$ | $-0.045 \pm 0.005$ | $0.012 \pm 0.003$ |
| V | $-0.045 \pm 0.009$ | $-0.004 \pm 0.002$ | — | — | $-0.044 \pm 0.005$ | $0.006 \pm 0.002$ |
| Cr | $-0.076 \pm 0.009$ | $-0.008 \pm 0.004$ | $-0.060 \pm 0.040$ | $0.002 \pm 0.010$ | $-0.053 \pm 0.005$ | $0.018 \pm 0.003$ |
| Mn | $-0.045 \pm 0.020$ | $-0.009 \pm 0.002$ | $-0.066 \pm 0.040$ | $-0.012 \pm 0.006$ | $-0.061 \pm 0.007$ | $0.005 \pm 0.002$ |
| Co | $-0.057 \pm 0.016$ | $-0.001 \pm 0.004$ | $-0.082 \pm 0.040$ | $-0.050 \pm 0.010$ | $-0.059 \pm 0.006$ | $0.003 \pm 0.004$ |
| Ni | $-0.078 \pm 0.011$ | $-0.004 \pm 0.002$ | $-0.062 \pm 0.040$ | $-0.009 \pm 0.006$ | $-0.053 \pm 0.005$ | $-0.003 \pm 0.002$ |
| Cu | $-0.071 \pm 0.009$ | $0.006 \pm 0.004$ | — | — | $-0.056 \pm 0.010$ | $0.002 \pm 0.003$ |
| Zn | $-0.053 \pm 0.016$ | $0.021 \pm 0.003$ | — | — | $-0.046 \pm 0.004$ | $0.012 \pm 0.002$ |
| Ce | $-0.047 \pm 0.018$ | $0.015 \pm 0.001$ | $-0.019 \pm 0.040$ | $0.074 \pm 0.020$ | $-0.037 \pm 0.004$ | $0.014 \pm 0.003$ |
| Nd | $-0.027 \pm 0.006$ | $0.013 \pm 0.003$ | $-0.092 \pm 0.040$ | $-0.044 \pm 0.010$ | $-0.015 \pm 0.003$ | $0.045 \pm 0.006$ |
| Yb | $-0.051 \pm 0.010$ | $0.029 \pm 0.002$ | — | — | — | — |

Table 1: Comparison of Galactic abundance gradients (slopes in dex/kpc) for [X/H], [X/Fe] vs. $R_{gc}$ from this work, Otto et al. (2026), and Magrini et al. (2023).

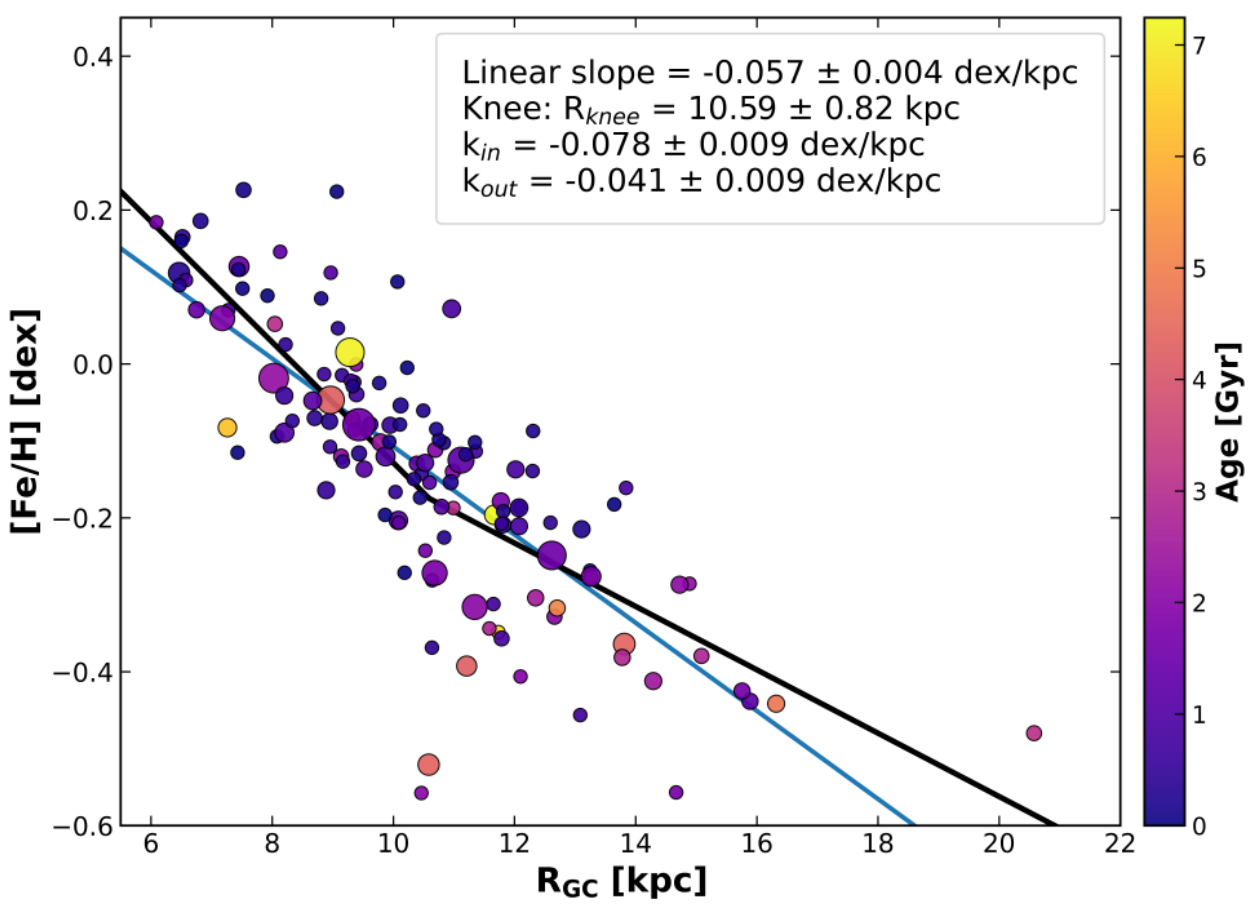

Fig. 4: [Fe/H] vs. $R_{gc}$ for our clusters. Each marker represents one cluster. Marker colour indicates the adopted cluster age according to the colour bar, while marker size scales with the number of analysed member stars (1-76 stars). The blue line shows the best-fit linear relation, while the black piecewise line shows the bilinear fit with a knee at $R_{knee}$ = 10.59 ± 0.82 kpc. The derived linear slope, knee location, and inner and outer bilinear slopes are listed in the legend.

radial selection, and age-dependent cluster survival.

Johnson et al. (2025), analysing APOGEE field giants, report all-star gradients of d[O/H]/d$R_{gc}$ = -0.062 ± 0.001 dex/kpc and d[Fe/H]/d$R_{gc}$ = -0.070 ± 0.003 dex/kpc. These measurements are not directly equivalent to those obtained here because the two studies differ in tracer population, age distribution, radial and vertical selection, abundance-analysis method, and statistical treatment. In particular, the present OC sample is dominated by relatively young systems, with a median age of approximately 0.83 Gyr and more than half of the clusters younger than 1 Gyr, whereas the APOGEE field-giant sample spans a much broader and, on average, older stellar population. The OCs for which high-quality abundance measurements are available may also represent a selected subset of the surviving OC population, owing to the combined effects of cluster disruption, detectability, membership selection, and spectroscopic targeting. Although field-star ages and distances have improved substantially through asteroseismology, data-driven methods, and Gaia astrometry, they remain comparatively uncertain and heterogeneous at the level of individual stars. OCs instead provide more tightly constrained ages and distances because these quantities can be inferred jointly from multiple coeval and chemically related members. The difference between our global OC metallicity slope and the field-star slope reported by Johnson et al. (2025) is approximately 0.013 dex/kpc, which is larger than the formal statistical uncertainty obtained by combining the quoted errors. This should not, however, be interpreted as a direct disagreement, because the formal uncertainties do not account for the substantial differences in population selection and systematic method-ology between the studies.

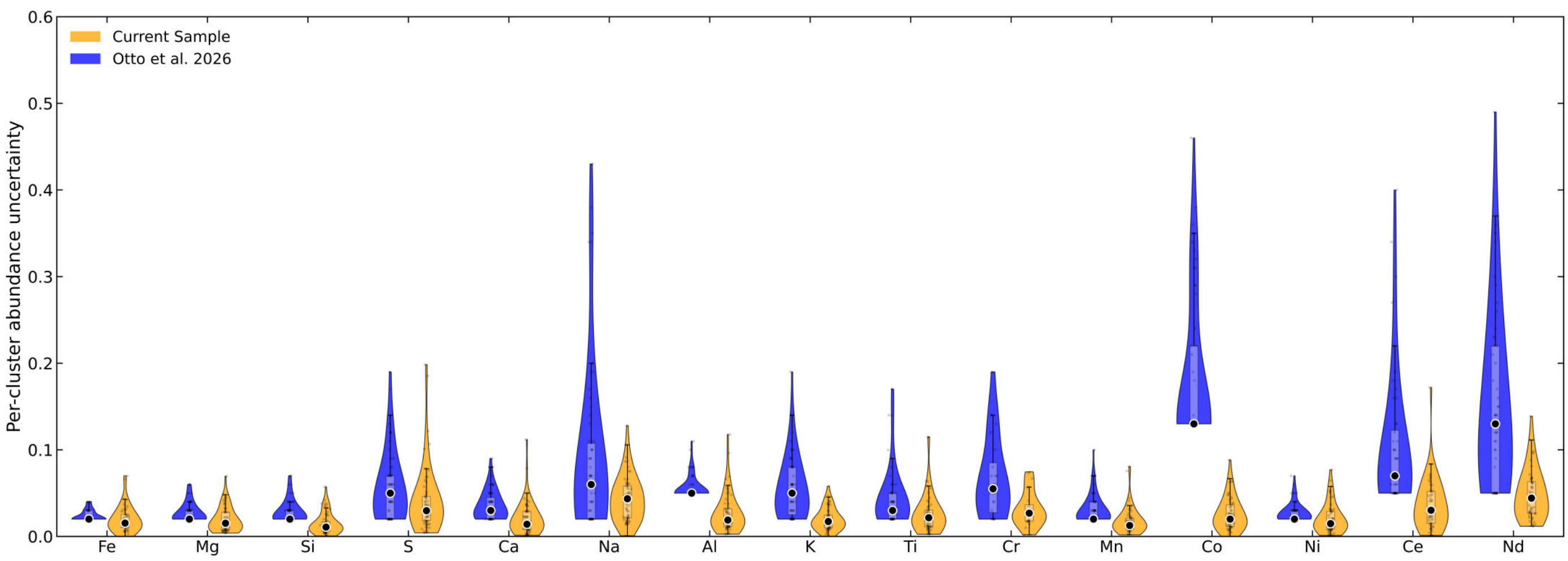

Fig. 5: Violin distributions of the reported per-cluster abundance uncertainties for the individual elements in the present sample (orange) and the SDSS-V Milky Way Mapper sample of Otto et al. (2026) (blue). For the present sample, each value is the propagated formal uncertainty on an unweighted mean cluster abundance, calculated as $\sqrt{\sum_{i=1}^{N}\sigma_i^2}/N$, where $\sigma_i$ is the formal uncertainty of an individual stellar measurement. For Otto et al. (2026), the plotted values are the quoted per-cluster uncertainties from that analysis. The uncertainty definitions may therefore not be strictly identical between the two studies. The violins show the distributions of reported cluster-mean uncertainties and do not represent the intrinsic star-to-star abundance dispersion within individual clusters. Lower and narrower violins indicate smaller reported per-cluster uncertainties.

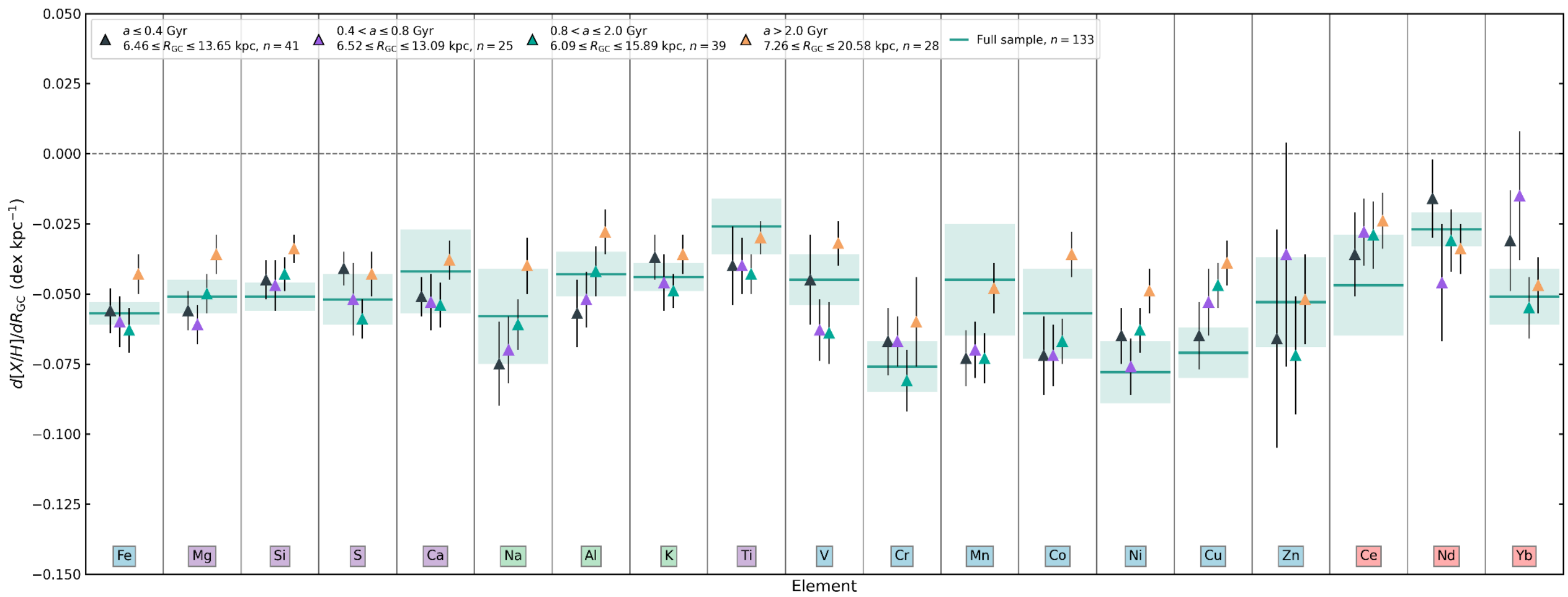

Fig. 6: The age dependence of the radial abundance gradients, $d[X/H]/dR_{gc}$, is shown for the OC sample. Coloured triangles show the slopes measured in four age bins: ≤0.4 Gyr, 0.4-0.8 Gyr, 0.8-2 Gyr and >2 Gyr. Vertical error bars represent the formal fit uncertainties. The horizontal teal line represents the slope calculated from the complete sample, while the shaded band indicates the uncertainties associated with the overall fit. Element symbols are displayed along the bottom. The gradients exhibit an age dependence, but not a strictly monotonic trend for all species.

A central result of Johnson et al. (2025) is that stellar populations younger than approximately 9 Gyr occupy broadly similar radial metallicity relations and do not show the strong monotonic decline in metallicity normalisation with age predicted by many classical GCE calculations, despite non-monotonic variations in the fitted slopes. They interpret this behaviour within an equilibrium framework in which the gas-phase metallicity approaches a radius-dependent equilibrium value relatively early in the evolution of the Galactic disc. In this interpretation, the radial abundance gradient is regulated by radial variations in star formation, gas accretion, and metal loss through outflows, rather than being determined solely by inside-out disc growth. Our OC results are qualitatively compatible with the absence of a simple monotonic evolution of the radial gradients. We do not observe a uniform progression from steep to shallow, or from shallow to steep, with increasing cluster age across all elements. Nevertheless, the tendency for some of the youngest cluster bins to show steeper gradients and for some of the oldest bins to show shallower gradients means that the present data do not demonstrate a strictly age-invariant radial abundance structure. The interpretation is limited by the restricted age coverage of the OC sample, the small number of old clusters, the correlation between cluster age and $R_{gc}$, and the possible effects of radial migration and age-dependent cluster disruption. The present results therefore neither confirm nor exclude the early-equilibrium interpretation of Johnson et al. (2025). They instead support the weaker conclusion that the radial chemical structure of the Galactic disc has not evolved in a simple, uniform, or strictly monotonic manner.

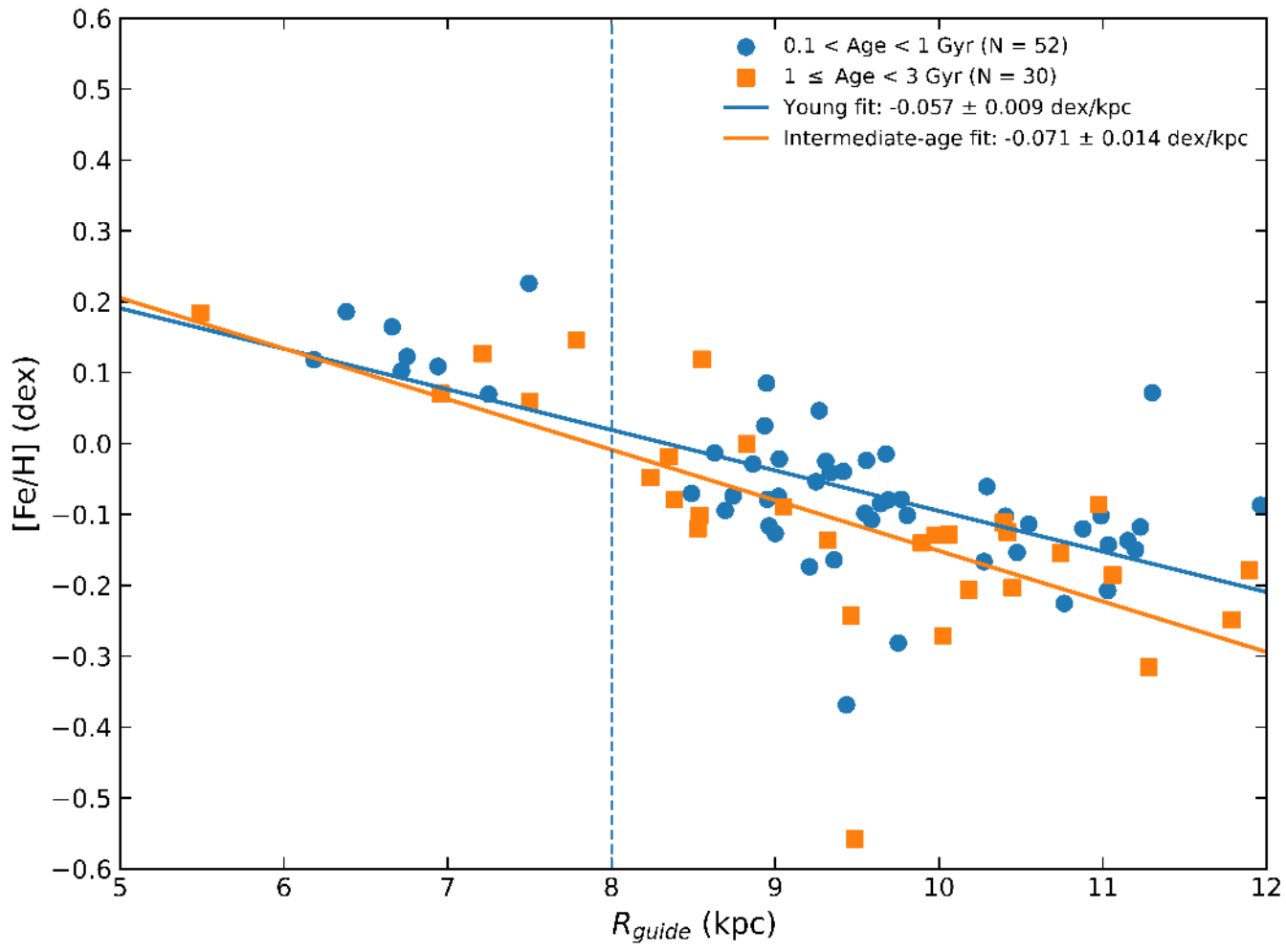

Fig. 7: Mean cluster [Fe/H] as a function of orbital guiding radius, $R_{guide}$, for the two age groups adopted in the comparison with Palla et al. (2024). The young group includes clusters with $0.1 < \text{Age} < 1$ Gyr, while the intermediate-age group includes clusters with $1 \leq \text{Age} < 3$ Gyr. Both samples are restricted to $R_{guide} < 12$ kpc. The solid lines show the linear fits to the two age groups, and the vertical dashed line marks $R_{guide} = 8$ kpc.

Palla et al. (2024) reported that, at a fixed orbital guiding radius, OCs younger than 1 Gyr are more metal-poor than clusters aged between 1 and 3 Gyr. By comparing this age-dependent displacement of the radial metallicity relation with two- and three-infall GCE models, they interpreted it as evidence for a recent gas-accretion episode that diluted the metal content of the disc ISM. We tested for the same observational signature in the present MWM sample using age and radial selections chosen to reproduce their comparison as closely as possible. We adopted $0.1 < \text{Age} < 1$ Gyr for the young population, $1 \leq \text{Age} < 3$ Gyr for the intermediate-age population, and restricted both samples to $R_{guide} < 12$ kpc. Following Palla et al. (2024), the guiding radius was defined as $R_{guide} = (R_{peri} + R_{apo})/2$. For each cluster, the median RA, DEC, proper motions, and radial velocity of the accepted member stars were combined with the adopted cluster distance to define the cluster phase-space coordinates. The cluster orbits were integrated for 5 Gyr using `galpy` and the `MWPotential2014` Galactic potential. We adopted $R_{\odot} = 8.3$ kpc, a circular velocity of 220 km/s, a solar height above the Galactic plane of 0.0208 kpc, and a solar peculiar velocity of (U, V, W) = (11.1, 12.24, 7.25) km/s. The pericentric and apocentric radii obtained from the integrated orbits were then used to calculate $R_{guide}$.

The adopted age and radial selections resulted in 52 young clusters and 30 intermediate-age clusters. The fitted radial metallicity gradients are -0.057 ± 0.009 dex/kpc for the young clusters and -0.071 ± 0.014 dex/kpc for the intermediate-age clusters, as shown in Fig. 7. The intermediate-age relation is therefore nominally steeper, in qualitative agreement with Palla et al. (2024) and Magrini et al. (2023). In particular, Palla et al. (2024) obtained slopes of approximately -0.050 and -0.093 dex/kpc for their young and intermediate-age samples, respectively, and -0.063 and -0.086 dex/kpc for their restricted sample. However, this nominal ordering depends on the adopted age intervals and is not recovered when the alternative age bins used in our main analysis are considered. Moreover, defining the slope difference as $\Delta\nabla = \nabla_{young} - \nabla_{intermediate}$, bootstrap resampling gives a 95% confidence interval of -0.040 to 0.010 dex/kpc. Because this interval includes zero, the difference between the two gradients is not statistically significant.

This nominal ordering differs from the general behaviour found in our main age-binned analysis, in which the oldest clusters exhibit shallower gradients than the youngest clusters. However, the two analyses use different age definitions and radial coverage. The comparison with Palla et al. (2024) adopts age intervals of 0.1-1 and 1-3 Gyr, excludes clusters older than 3 Gyr, and is restricted to $R_{guide} < 12$ kpc. The intermediate-age sample is also dominated by clusters between 1 and 2 Gyr and is therefore not equivalent to the oldest bin in our main analysis. The strongest sensitivity arises from the radial restriction. When the $R_{guide} < 12$ kpc cut is removed, the young and intermediate-age gradients become -0.053 ± 0.006 and -0.057 ± 0.007 dex/kpc, respectively, and are therefore nearly identical within their uncertainties. The apparent difference from the main age-binned analysis is consequently associated primarily with the adopted radial coverage and age definitions, rather than indicating a robust reversal of the age dependence.

At $R_{guide} = 8$ kpc, the fitted metallicities are 0.019 ± 0.017 dex for the young clusters and -0.009 ± 0.028 dex for the intermediate-age clusters. Defining the metallicity displacement as $\Delta[Fe/H] = [Fe/H]_{young} - [Fe/H]_{intermediate}$, the corresponding difference is 0.028 ± 0.028 dex, with a bootstrap 95% confidence interval of -0.023 to 0.086 dex and a permutation-test result of $p = 0.28$. We therefore find no statistically significant metallicity displacement between the two populations. Activity-related abundance systematics may also depend on the wavelength range and analysis methodology, and the use of NIR rather than optical spectra could contribute to differences between studies. Nevertheless, we find no evidence that the youngest clusters in the MWM sample are systematically displaced towards artificially low metallicities. Although the fitted gradients show the same nominal ordering as those of Palla et al. (2024), the MWM sample does not reproduce the lower metallicity of the youngest OCs reported in that study and therefore provides no independent evidence for the proposed recent-dilution signature.

This comparison highlights the importance of distinguishing between chemical and chemo-dynamical evolution. In classical inside-out disc-formation models, such as those presented by Chiappini (2009), negative radial metallicity gradients arise because the inner disc forms and enriches more rapidly than the outer regions. Within this framework, the negative [Fe/H] gradient measured for the OC sample is expected, while the observed flattening beyond $R_{gc} \simeq$ 10-11 kpc indicates that the outer-disc metallicity distribution cannot be described by a straightforward extrapolation of the inner-disc gradient. The flattening may instead reflect radial variations in star-formation efficiency, gas-accretion history, radial gas flows, or the relative contribution of different cluster populations at large $R_{gc}$.

The age dependence of the OC gradients should be interpreted with caution because the present-day cluster distribution does not directly trace the ISM gradient at the time of cluster formation. Chemo-dynamical models (e.g., Minchev et al. 2013; Minchev et al. 2014) demonstrate that radial migration, disc heating, flaring, and the evolving radial and vertical distributions of stellar populations can alter present-day abundance gradients. Consequently, the steeper gradients observed for some of the youngest cluster bins and the generally flatter gradients found for some of the oldest bins may reflect not only temporal evolution of the ISM gradient, but also radial migration, cluster-survival biases, and differences in the radial coverage of the age bins. The tendency for some old-cluster populations to exhibit flatter gradients is qualitatively compatible with longer dynamical redistribution times and with the preferential survival of old clusters on particular orbits. However, the non-monotonic and element-dependent behaviour shown in Fig. 6 does not provide a unique signature of radial migration.

## 5. Conclusions and summary

We analysed a large, homogeneous sample of open cluster giants using APOGEE spectra from MWM DR19. The selected dataset comprises 655 stars in 133 clusters with ages of approximately 0.2-7.3 Gyr and distances of approximately 5-21 kpc from the Galactic Centre. We derived stellar parameters and abundances for 18 elements through spectrum fitting with PySME, using MARCS atmospheres and NLTE synthesis for several elements. This approach yields precise stellar parameters and a revised abundance scale relative to the MWM pipeline analysis and the work of Otto et al. (2026). The propagated formal uncertainties on the mean cluster abundances are typically 0.01-0.10 dex.

Of the measured species, all [X/H] gradients are negative and generally fall within a comparable range of -0.03 to -0.08 dex/kpc. In contrast, most [X/Fe] gradients are weak, nearly flat, or only mildly positive. The $\alpha$-elements broadly follow the metallicity trend, with [Mg/H], [Si/H], [S/H], [Ca/H] and [Ti/H] all declining with increasing radius. Meanwhile, the corresponding differential ratios relative to Fe are much less structured, remaining close to constant or showing slight positive slopes. The Fe-peak elements exhibit a similar pattern: their [X/H] relations are consistently negative, whereas their [X/Fe] slopes tend to be close to zero. In contrast, several odd-Z and neutron-capture elements exhibit more distinct element-specific behaviour. Notably, K, Al, Ce, and Nd show mild enhancement relative to Fe towards the outer disk, while [Na/Fe] remains approximately flat. The trends in [Ce/Fe] and [Nd/Fe] are notable in that they both increase gently with $R_{gc}$, even though [Ce/H] and [Nd/H] decrease outward. Compared with Otto et al. (2026), the revised analysis yields a tighter abundance scale overall and, for several chemically important species, less extreme [X/Fe] gradients.

The present analysis reveals that the radial abundance gradients exhibit variation with the age of the cluster. However, this variation is neither strictly monotonic nor uniform across all elements. For the majority of elements, the [X/H] gradients within different age bins remain negative, aligning with the inside-out formation scenario of the Galactic disk–wherein the inner disk attains a higher chemical enrichment than the outer regions (e.g. Chiappini et al. 1997; Boissier & Prantzos 1999). The observed deviations from a simple monotonic age dependence likely arise from the interplay of several factors: element-specific enrichment processes, radial migration, sample selection biases, and the limited number of clusters represented in certain age bins. Nevertheless, several elements, including Fe, Mg, Ca, Na, Mn, Co, and Ni, exhibit comparatively steep gradients in the youngest age bin, whereas the oldest clusters often exhibit flatter slopes. The $\alpha$-elements tend to be more tightly grouped between age bins, whereas the odd-Z, Fe-peak and neutron-capture elements exhibit greater differences between elements. Notably, Ce, Nd, and Yb do not follow a straightforward age sequence; some age bins exhibit weak, nearly flat, or even slightly positive gradients within the margin of error. This behavior differs from that observed by Magrini et al. (2023), who found flatter gradients for the youngest Gaia-ESO clusters. In the present sample, many young clusters populate the radial range over which the current abundance gradient is well defined, meaning that the youngest bin retains a relatively steep present-day gradient. Conversely, older clusters are more likely to have been affected by dynamical evolution and radial migration, processes that can disperse clusters from their birth radii and weaken the observed correlation between abundance and present-day $R_{gc}$. This interpretation is consistent with chemically-dynamical models, such as those presented in Minchev et al. (2014), in which radial migration and the time-dependent growth of the disc modify the observed age dependence of radial gradients. Using the orbital guiding radii and age selections adopted by Palla et al. (2024), we find no statistically significant metallicity difference between clusters younger than 1 Gyr and those aged 1-3 Gyr, and therefore find no independent evidence in the MWM sample for the proposed recent-dilution signature.

In conclusion, our element-by-element analysis of APOGEE MWM DR19 OC data reveals smooth radial abundance gradients across the Galactic disc, together with modest age-dependent variations in these gradients. The carefully vetted sample and the generally small propagated formal uncertainties on the cluster mean abundances help define subtle abundance trends, providing a detailed view of the chemical structure of the Galactic disc.

*Acknowledgements.* We would like to thank the anonymous referee for helpful suggestions that improved this paper in many ways.
S.B.S. acknowledges support from the Foundation for the Promotion of the Development of Malmö University.
H.J. acknowledges support from the Swedish Research Council, VR (grant 2024-04989).
VD acknowledges support from the INAF Minigrant 2024 MUGS.
This work has made use of the VALD database, operated at Uppsala University, the Institute of Astronomy RAS in Moscow, and the University of Vienna.

## Data availability

All machine-readable data underlying this work are available in electronic form at the CDS. The CDS package includes: (i) the full list of observed stars and basic properties for each OC (Appendix A); (ii) the complete atomic and molecular line list used in the analysis (Appendix B); (iii) derived stellar parameters with formal PySME uncertainties and (iv) per-star abundance ratios [X/Fe] with uncertainties and cluster mean values documenting sample completeness (Appendix D). All files include column descriptions and are provided in machine-readable ASCII format.

**All appendix tables are available electronically via the CDS.**

## Appendix A: Spatial arrangement of the cluster sample and a sample overview

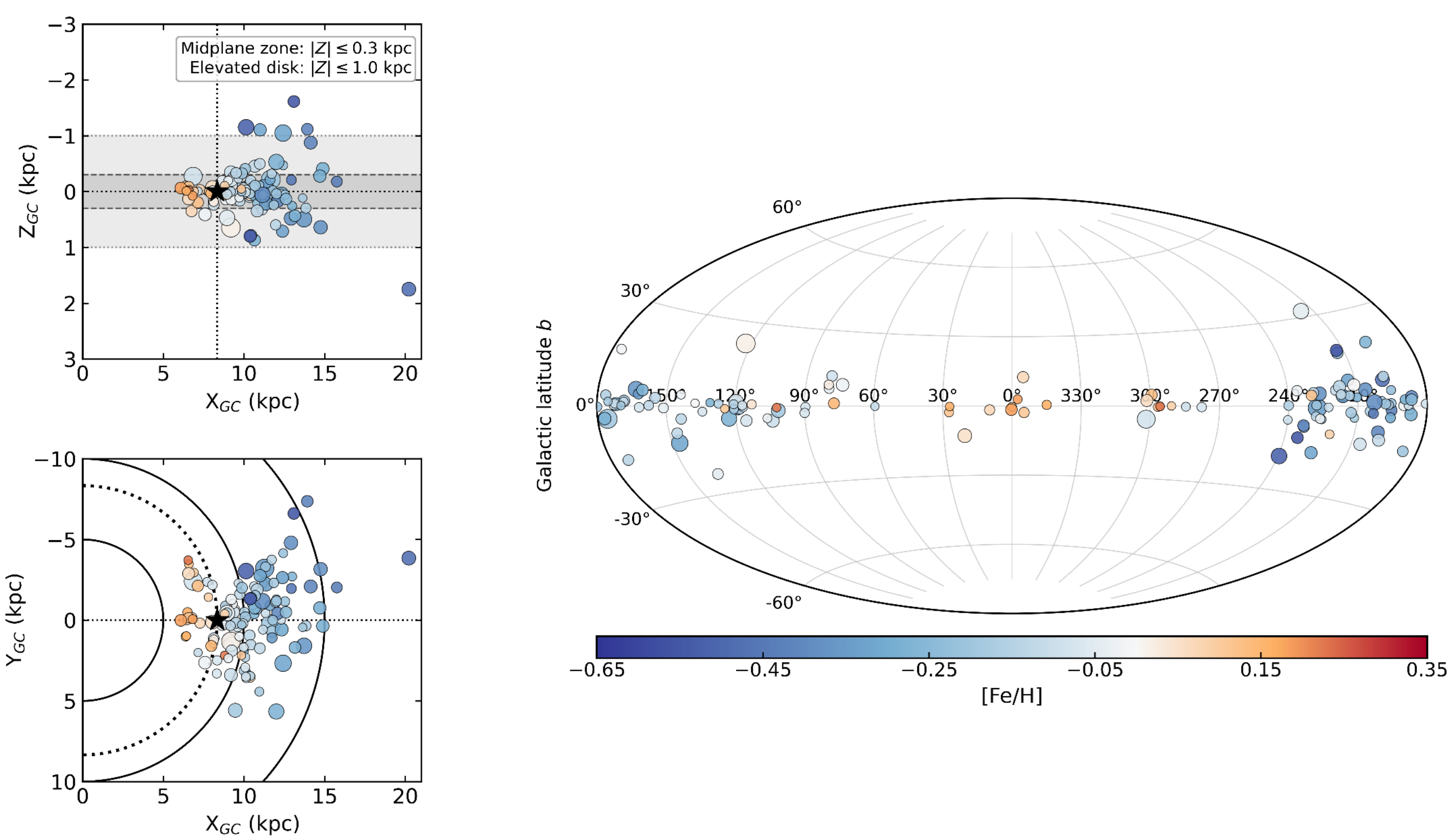

Fig. A.1: The spatial distribution of the OC sample is shown in Galactocentric and Galactic coordinates. The left-hand panels show the cluster positions in the $X_{GC}$–$Z_{GC}$ plane (top) and the $X_{GC}$–$Y_{GC}$ plane (bottom). The black star marks the position of the Sun at $R_{\odot} = 8.34$ kpc; the vertical dotted line indicates the solar Galactocentric radius; and the shaded region in the top panel highlights the disk plane within $|Z| \lesssim 0.3$ kpc. The right panel shows the corresponding distribution in Galactic longitude and latitude. The symbols are color-coded by mean cluster metallicity, with symbol size increasing with cluster age.

Table A.1: General information for the stars analyzed in each cluster, including the Gaia DR3 source identifier, equatorial coordinates, and the Gaia $G$-band magnitude. Cluster ages, $A_V$, and $R_{gc}$ are adopted from Cantat-Gaudin et al. (2020). Clusters are listed in ascending order of age.

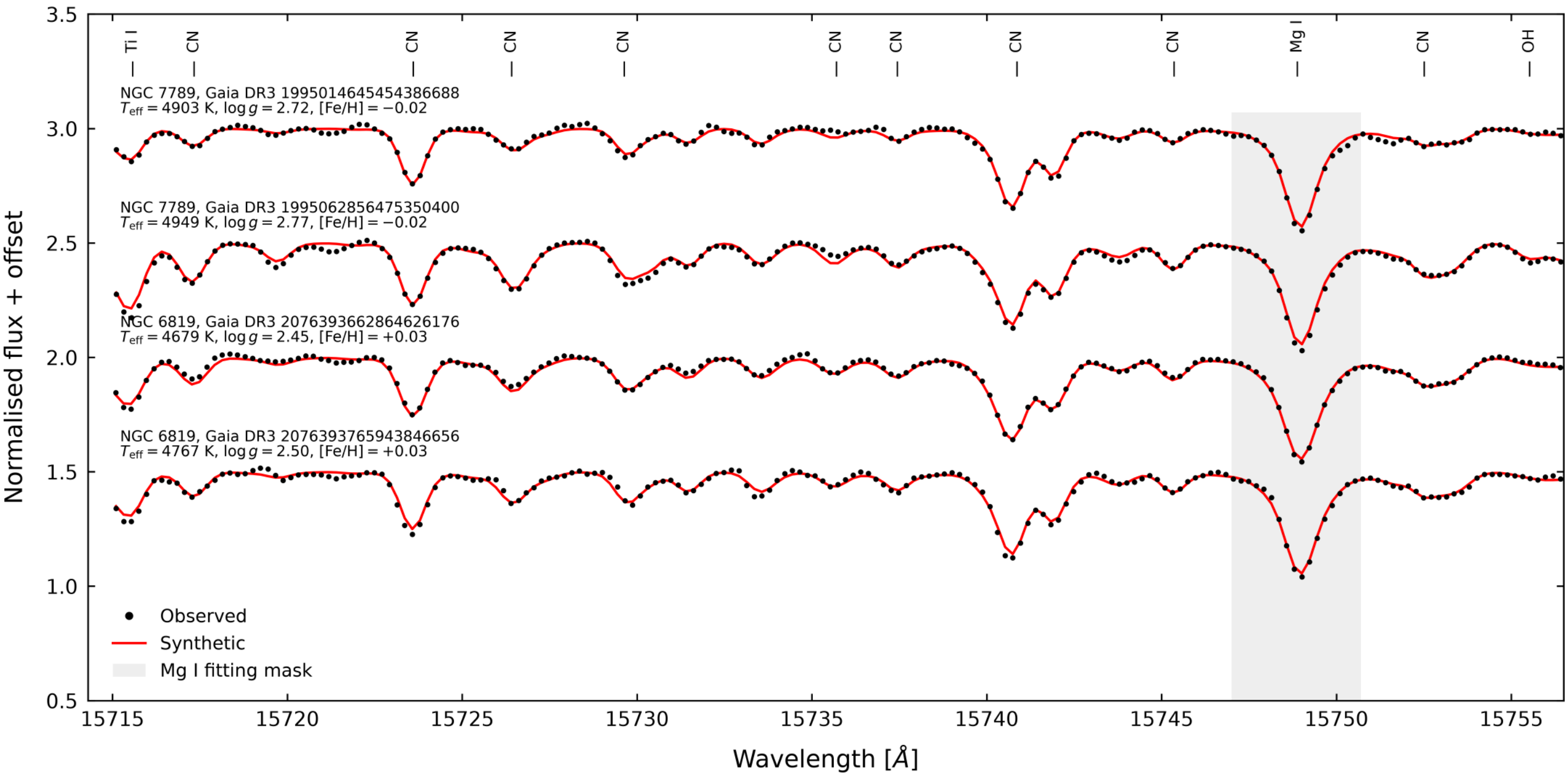

Fig. A.2: Representative examples of accepted spectral fits. The black points show the observed spectra, while the red curves show the corresponding best-fitting synthetic spectra. The spectra are vertically offset for clarity. The cluster name, GAIA-DR3 ID, $T_{eff}$, $\log g$, and [Fe/H] are given for each star. Some atomic and molecular transitions included in the adopted synthesis line list are identified at the top of the panel. The shaded region marks the fitting mask around the Mg I line at 15748.886 Å.

## Appendix B: Atomic and molecular data

Table B.1: Line data adopted for the abundance analysis, including molecular lines of OH, CO, and CN, and atomic lines of Fe I, Mg I, Si I, Ti I/II, S I, Ca I, Na I, Al I, K I, V I, Cr I, Mn I, Co I, Ni I, Cu I, Zn I, Ce II, Nd II, and Yb I. The $\log(gf)$-values are from Brooke et al. (2016) (OH), Li et al. (2015) (CO), and Sneden et al. (2014) (CN). All remaining transitions are from VALD (Piskunov et al. 1995; Kupka et al. 2000; Ryabchikova et al. 2015) but updated with astrophysical calibrations based on solar spectra (Montelius et al. 2022; Nandakumar et al. 2023a,b, 2024b).

## Appendix C: Kiel diagram and parameter comparison with OCCAM DR19

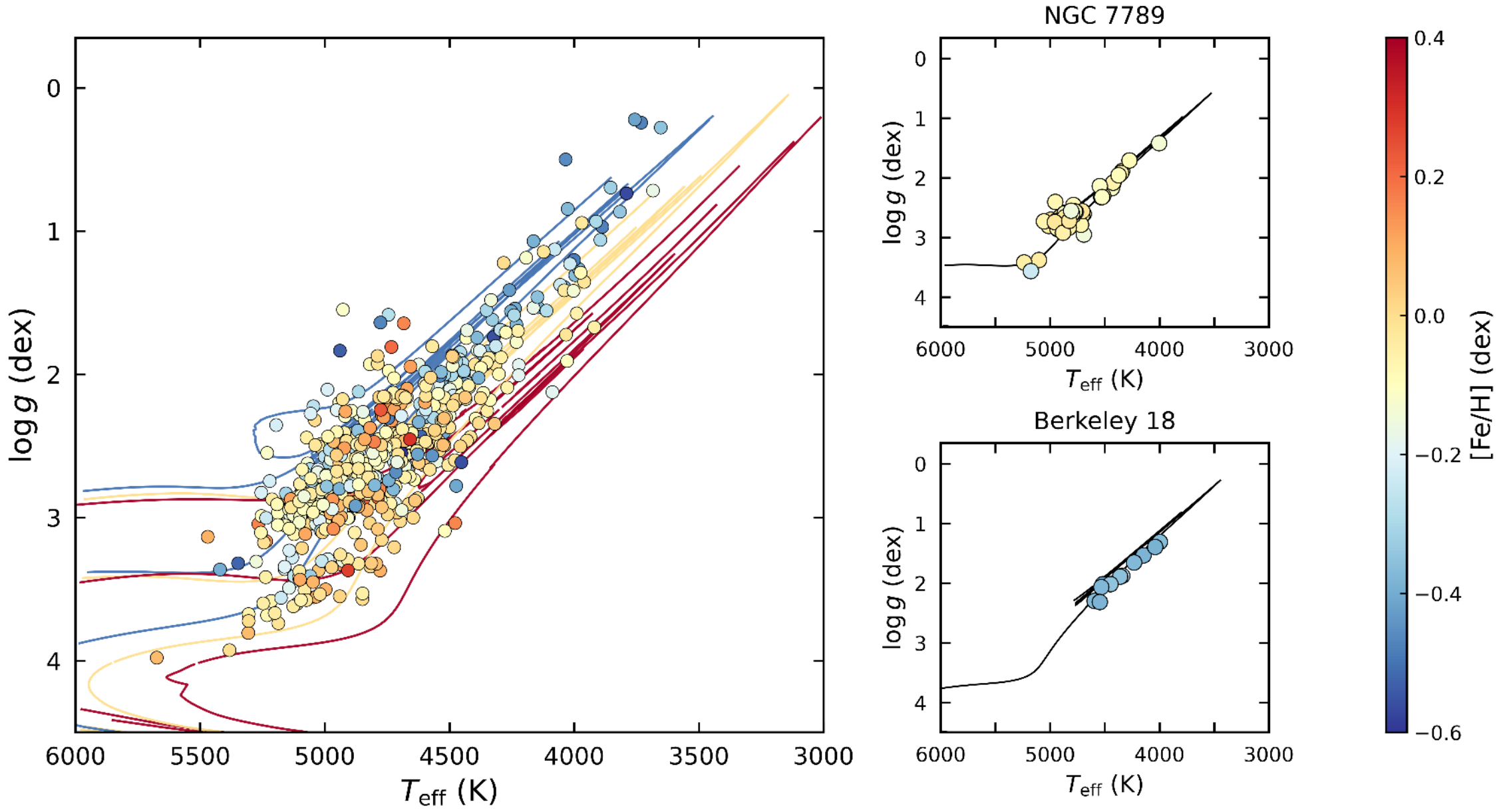

Fig. C.1: The main panel shows the full stellar sample in the Kiel diagram, colour-coded by [Fe/H]. The MIST isochrone grid comprises three representative cluster ages spanning the age range of the sample, each evaluated at three metallicities, [Fe/H] = -0.50, 0.00, and 0.50 dex, giving a total of nine isochrones. The isochrones are coloured according to metallicity, and evolutionary phases 0-4 are included. The right-hand panels show individual comparisons for NGC 7789 (top) and Berkeley 18 (bottom). The black curves show MIST isochrones computed using the cluster parameters adopted in this work: age = 1.55 Gyr and [Fe/H] = -0.08 dex for NGC 7789, and age = 4.37 Gyr and [Fe/H] = -0.36 dex for Berkeley 18.

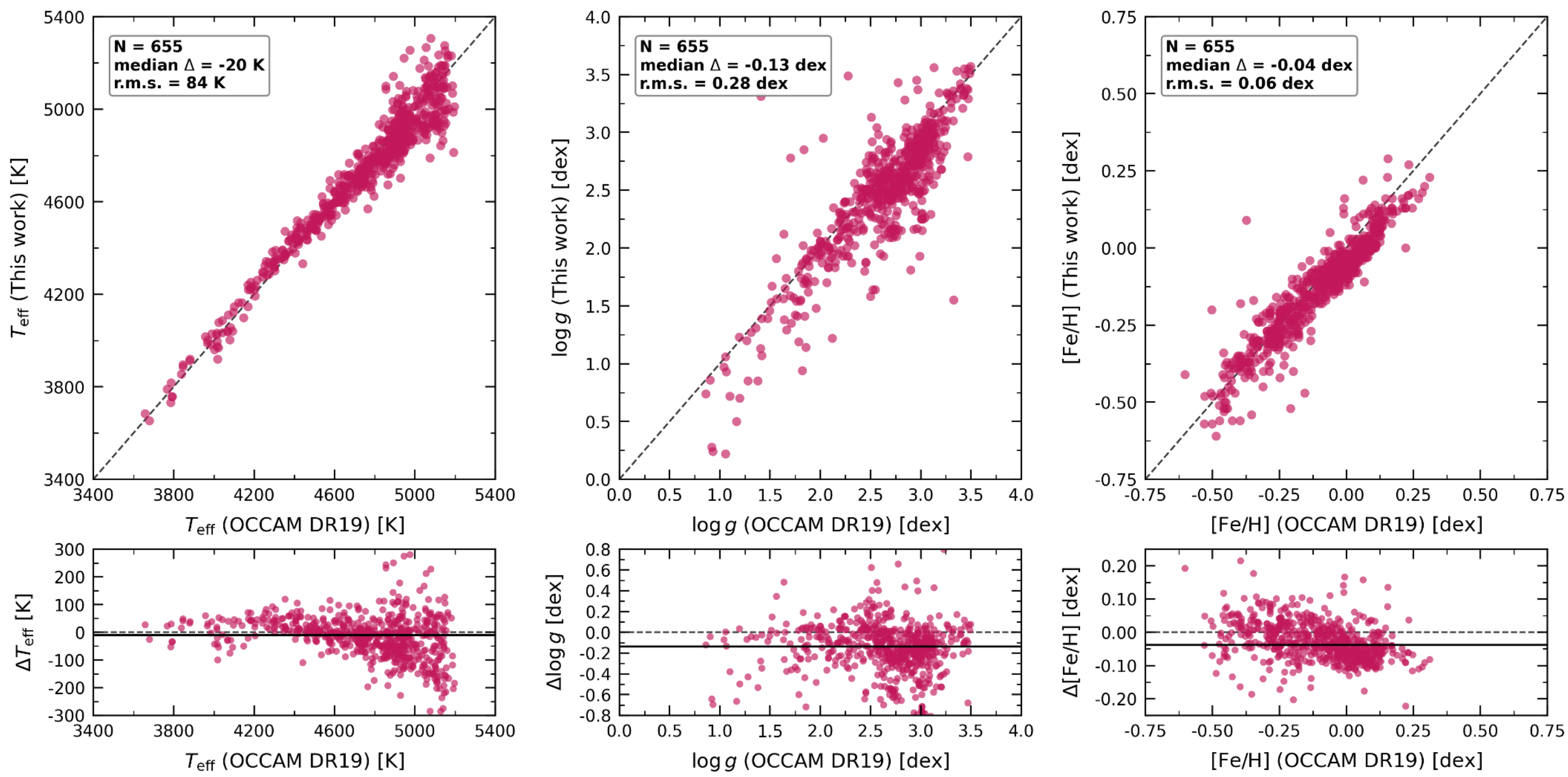

Fig. C.2: The full sample contains 655 stars, which were used to compare with OCCAM DR19 catalogue parameters. The largest systematic difference is found in $\log g$, where the photometrically anchored values adopted in the present work are lower than the OCCAM DR19 spectroscopic values by a median of -0.13 dex.

## Appendix D: Stellar parameters and Abundances

Table D.1: Derived stellar parameters and CNO abundances for the sample stars: $T_{eff}$, $\log g$, [Fe/H], $v_{mic}$, $v_{mac}$, [C/Fe], [N/Fe], and [O/Fe]. The quoted uncertainties correspond to the formal PySME fit errors.
Table D.2: Abundance ratios [Mg/Fe], [Si/Fe], [S/Fe], [Ca/Fe], [Na/Fe], [Al/Fe], [K/Fe], [Ti/Fe] and [V/Fe] for the individual sample stars.
Table D.3: Abundance ratios [Cr/Fe], [Mn/Fe], [Co/Fe], [Ni/Fe], [Cu/Fe], [Zn/Fe], [Ce/Fe], [Nd/Fe], and [Yb/Fe] for the individual sample stars.
Table D.4: Cluster mean values of [Mg/Fe], [Si/Fe], [S/Fe], [Ca/Fe], [Na/Fe], [Al/Fe], and [K/Fe] from our sample.
Table D.5: Cluster mean values of [Ti/Fe], [V/Fe], [Cr/Fe], [Mn/Fe], [Co/Fe], and [Ni/Fe] from our sample.
Table D.6: Cluster mean values of [Cu/Fe], [Zn/Fe], [Ce/Fe], [Nd/Fe], and [Yb/Fe] from our sample.
For Tables D.4–D.6, the quoted $\pm$ values represent the propagated formal uncertainty on the mean cluster abundance. For a cluster with $N$ valid stellar measurements, this is calculated as $\sqrt{\sum_{i=1}^{N} \sigma_i^2}/N$, where $\sigma_i$ is the formal uncertainty of the individual stellar abundance measurement. For a cluster represented by only one valid measurement, the quoted uncertainty is the formal uncertainty of that measurement.